\documentstyle{mn}  

\input epsf
  
\begin{document}  
  
\title{THE MORPHOLOGY OF HII GALAXIES}\label{chap:morph}  
  
\author[Telles, Melnick \& Terlevich] {Eduardo Telles$^{1,2}$\thanks{present address:
Instituto Astron\^omico e Geof\'{\i}sico - USP,
Caixa Postal 9638,
01065-970 - S\~ao Paulo - BRASIL}
\and
Jorge Melnick$^3$ \and Roberto  
Terlevich$^2$ \\ 1. Institute of Astronomy, Madingley Road, Cambridge  
CB3 0HA, U.K. \\  2. Royal Greenwich  
Observatory, Madingley Road, Cambridge CB3 0EZ, U.K.\\
3. European Southern Observatory, La Silla, Chile \\  
etelles@cosmos.iagusp.usp.br, jmelnick@eso.org \& rjt@ast.cam.ac.uk}

\date{accepted January 7, 1997}

\maketitle  
  
\def\cao{\c{c}\~ao~}
\def\CAO{\c{C}\~AO~}
\def\coes{\c{c}\~oes~}
\def\COES{\c{C}\~OES~}
%
%
\def\a4{\hsize 15.cm \vsize 23.cm}
\def\skipline{\vskip 10.1pt}
\def\double{\baselineskip 18pt \lineskip 10pt}     
\def\semidouble{\baselineskip 13pt \lineskip 6pt}
\def\single{\baselineskip 12pt \lineskip 1pt}
\def\supersingle{\baselineskip 10pt \lineskip 1pt}
\def\newpage{\vfill\eject}
%
%
\def\Diaz{\hbox{D\'\i az}}
\def\Vilchez{\hbox{V\'\i lchez}}
\def\neq{\mathrel{\not=}}
\def\twiddles{\hbox{$\sim $}}
\def\varomega{\varpi}
%
%
\def\Angeles{\hbox{Angeles I. D\'\i az}}
\def\Pepe{\hbox{Jos\'e M. V\'\i lchez}}
\def\Elena{\hbox{Elena Terlevich}}
\def\Bernard{\hbox{Bernard E.J. Pagel}}
\def\Mike{\hbox{Michael G. Edmunds}}
%
%
\def\ang{\thinspace\hbox{\AA}}
\def\km{\thinspace\hbox{km}}
\def\Mpc{\thinspace\hbox{Mpc}}
\def\kpc{\thinspace\hbox{kpc}}
\def\kmsec{\thinspace\hbox{$\hbox{km}\thinspace\hbox{s}^{-1}$}}
\def\kmsecmeg{\thinspace\kmsec\Mpc$^{-1}$}
\def\ergsec{\thinspace\hbox{$\hbox{erg}\thinspace\hbox{s}^{-1}$}}
\def\ergsqcmsec{\thinspace\hbox{erg}\sqcm\sec$^{-1}$}
\def\sqcm{\thinspace\hbox{$\hbox{cm}^{2}$}}
\def\cm3{\thinspace\hbox{$\hbox{cm}^{3}$}}
\def\cm2{\thinspace\hbox{$\hbox{cm}^{2}$}}
\def\persqcm{\thinspace\hbox{$\hbox{cm}^{-2}$}}
\def\percucm{\thinspace\hbox{$\hbox{cm}^{-3}$}}
\def\kev{\thinspace\hbox{keV}}
\def\sec{\thinspace\hbox{s}}
\def\ha{\hbox{$\hbox{H}_\alpha$}}
\def\hb{\hbox{$\hbox{H}_\beta$}}
\def\hg{\hbox{$\hbox{H}_\gamma$}}
\def\hd{\hbox{$\hbox{H}_\delta$}}
\def\lya{\hbox{$\hbox{Ly}_\alpha$}}
\def\kelvin{\thinspace\hbox{K}}
\def\dyncm2{\thinspace\hbox{$\hbox{dyn}\thinspace\hbox{cm}^{-2}$}}
\def\deg{\hbox{$^\circ$}}
\def\rstar{\thinspace\hbox{$\hbox{R}_*$}}
\def\vstar{\thinspace\hbox{$\hbox{V}_*$}}
\def\Zsun{\thinspace\hbox{$\hbox{Z}_{\odot}$}}
\def\msun{\thinspace\hbox{$\hbox{M}_{\odot}$}}
\def\rsun{\thinspace\hbox{$\hbox{R}_{\odot}$}}
\def\lsun{\thinspace\hbox{$\hbox{L}_{\odot}$}}
\def\gpar{\hbox{$g_{\parallel}$}}
\def\Ca{$\lambda\lambda$\thinspace\hbox{8498,8542,8662}~\AA}
\def\C2{$\lambda\lambda$\thinspace\hbox{8542,8662}~\AA}
\def\Mg{$\lambda$\thinspace\hbox{8807}~\AA}
\def\Fe{\hbox{$[Fe/H]$}}
\def\lg{\hbox{$log\thinspace\hbox{g}$}}
\def\lT{\hbox{$log\thinspace\hbox{T}_{eff}$}}
\def\Teff{\hbox{$T_{eff}$}}
%
%
\def\UAM {\ninesl {Depto. de F\'\i sica Te\'orica, C-XI, 
Univ. Aut\'onoma de Madrid, 28049 Madrid, Spain.}}
\def\RGO {\ninesl{Royal Greenwich Observatory, Madingley Rd.,
Cambridge, CB3 0EZ, U.K.}}
\def\IAC {\ninesl{Instituto de Astrof\'\i sica de Canarias, 
38200 La Laguna, Tenerife, Spain}}
\def\Cardiff {\ninesl{Department of Physics, University of Wales 
College of Cardiff, PO Box 913, Cardiff, CF1 3TH, U.K.}}
\def\Nordita {\ninesl{NORDITA, Blegdamsvej 17, DK-2100 K\obenhavn \O,
Denmark}}
%
\def\reference#1#2#3#4#5{\tenp\par\noindent\hangindent 3em
              #1, #2. {\tenpsl #3\/}, {\tenpb #4,}
\thinspace\hbox{#5}}
\def\refnum#1#2#3#4{\tenp\noindent #1, #2. {\sl #3\/}, {\bf #4}}
\def\refpress#1#2#3{\tenp\par\noindent\hangindent 3em
             #1, #2. {\tenpsl #3\/}, in press}
\def\refsub#1#2#3{\tenp\par\noindent\hangindent 3em
           #1, #2. {\tenpsl #3\/}, submitted}
\def\refprep#1#2#3{\tenp\par\noindent\hangindent 3em
           #1, #2. {\tenpsl #3\/}, in preparation}
\def\refacc#1#2#3{\tenp\par\noindent\hangindent 3em
           #1, #2. {\tenpsl #3\/}, accepted}
\def\refbook#1#2#3{\tenp\par\noindent\hangindent 3em
            #1, #2. {\tenpsl #3\/}}

\def\etal   {et\nobreak\ al.\ }
\def\etalp   {{\tenib et\nobreak\ al.\ }}
\def\etalr   {{\tenpsl et\nobreak\ al.\ }}
\def\aanda  {Astr.\ Astro\-phys.\nobreak\ }
\def\aandas {Astr.\ Astro\-phys.\nobreak\ Suppl.\nobreak\ }
\def\aandal {Astr.\ Astro\-phys.\nobreak\ Lett.\nobreak\ }
\def\aj     {Astron.\nobreak\ J.\nobreak\ }
\def\annrev {Ann.\ Rev.\ Astr.\ Astro\-phys.\nobreak\ }
\def\acta   {Acta Astron.\nobreak\ }
\def\apj    {Astro\-phys.\nobreak\ J.\nobreak\ }
\def\apjs   {Astro\-phys.\nobreak\ J.\ Suppl.\nobreak\ }
\def\apjl   {Astro\-phys.\nobreak\ J.\ Lett.\nobreak\ }
\def\apspsc {Astro\-phys.\nobreak\ Sp.\nobreak\ Sc.\nobreak\ }
\def\apjsupp{\apjs}
\def\aplett {Astro\-phys.\nobreak\ Lett.}
\def\commap {Comments\ Astrophys.\nobreak\ }
\def\ibvs   {Inf.\ Bull.\ var.\ Stars}
\def\mn     {Mon.\ Not.\ R.\ astr.\nobreak\ Soc.\nobreak\ }
\def\pasp   {Publ.\ astr.\ Soc.\ Pacif.\nobreak\ }
\def\pasj   {Publ.\ astr.\ Soc.\ Japan\nobreak\ }
\def\sovast {Soviet astr.}
\def\rmx    {Rev.\ Mex.\ Astr.\ Astrofis.\nobreak\ }
\def\jump   {\par\vskip 0.2cm\noindent\hangindent 3em}
%
%
\newcount\notenumber
\def\clearnotenumber{\notenumber=0}
\def\note{\advance\notenumber by 1
  \footnote{$^{\the\notenumber}$}}
\newcount\natrefreg
\def\clearnatref{\natrefreg=0}
\def\natref{\advance\natrefreg by 1 $^{\the\natrefreg}$}
\newcount\secreg
\def\clearsecreg{\secreg=0}
\newcount\subsecreg
\def\clearsubsecreg{\subsecreg=0}
\def\newsection#1{\skipline\skipline\par\smallbreak\noindent\advance\secreg by1
 \the\secreg . {\bf #1}\nobreak\par\clearsubsecreg}
\newcount\tablereg
\def\cleartable{\tablereg=0}
\def\nextable{\advance\tablereg by1\table}
\newcount\figreg
\def\clearfig{\figreg=0}
\def\fig{Fig.~\the\figreg}
\def\nextfig{\advance\figreg by1\fig}
%
%
\def\rsa{{\tenpsl Revised Shap\-ley-Ames Catalog of Bright Gal\-axies\/} Sandage
 \&\ Tammann (1981)}
\def\RC2{{\tenpsl Second Reference Catalogue of Bright Gal\-axies\/},
 de~Vaucouleurs, de~Vaucouleurs and Corwin (1976)}
\def\rc2{\RC2}
\def\Sin{\hbox{Sin}}
\def\Cos{\hbox{Cos}}
%
%
\def\deriv#1#2{\hbox{${{\displaystyle\hbox{d}#1}\over
{\displaystyle\hbox{d}#2}}$}}
\def\sderiv#1#2{\hbox{${{\displaystyle\hbox{d}^2#1}\over
{\displaystyle\hbox{d}#2^2}}$}}
\def\pderiv#1#2{\hbox{${{\displaystyle\partial#1}\over
{\displaystyle\partial#2}}$}}
\def\spderiv#1#2{\hbox{${{\displaystyle\partial^2#1}\over
{\displaystyle\partial#2^2}}$}}
\def\x10#1{\hbox{$\times\hbox{10}^{#1}$}}
\def\eex#1{\hbox{$\hbox{10}^{#1}$}}
\def\movements#1{\halign{\quad\it##\hfil&&\qquad\it##\hfil\cr#1\crcr}}
%
%
\def\integ{\int\limits}
%
%
\font\title=cmbx10 scaled 1440
\font\subt=cmbx10 scaled 1200
\font\tenp=cmr10
\font\poster=cmr10 scaled 1440
\font\tenpoint=cmr10 scaled 1200
\font\nine=cmr9
\font\ninei=cmti9
\font\nineb=cmbx9
\font\tenpib=cmbxti10
\font\tenib=cmbxti10 scaled 1200
\font\ninesl=cmsl9
\font\mit=cmmi10 scaled 1200
\font\tenpsl=cmsl10
\font\tenpb=cmbx10  
  
\begin{abstract}  

We present CCD images of a sample of 39 HII galaxies taken at the
Danish 1.54m telescope on La Silla.  The images are used to analyse
the morphology of these emission line dwarfs, and the structural
properties of the knots of star formation and of the underlying
galaxy.  The sizes of the starbursts are measured.  We propose a
morphological classification based on the presence or absence of signs
of tails, extensions, or distorted outer isophotes. This criterion
segregates the objects into two broad morphological types with
different physical properties: the more disturbed and extended (type
I) HII galaxies having larger luminosities and velocity dispersions
than the more compact and regular (type II) objects.  The relative
position of HII galaxies and of a sample of dwarf elliptical galaxies
in the [R -- $\sigma$] diagram support the hypothesis of a possible
evolutionary link between the two types of galaxy.

\end{abstract}


\section{Introduction}

HII galaxies are narrow emission line dwarf galaxies undergoing
violent star formation (Melnick, Terlevich \& Eggleton 1985) whose
spectroscopic properties are indistinguishable from extragalactic
giant HII regions in normal late type galaxies (e.g. 30 Dor in LMC,
NGC 604 in M33) (Sargent \& Searle 1970).  Their high rates of star
formation and low heavy element abundances imply that the star
formation history must be simple and episodic (i.e. few burst of short
duration followed by long quiescent periods).  A recent review on the
global properties of HII galaxies is given by Telles (1995, and
references therein). The possible links of HII galaxies with other
types of known dwarf galaxies have been discussed by Thuan (1983);
Loose \& Thuan (1985); Bothun \etal (1986); Kunth, Maurogordato \&
Vigroux (1988); Davies \& Phillipps (1988), Drinkwater \& Hardy
(1991).  However, no conclusive answer has been given to the questions
of what these systems will resemble when the present period of violent
star formation ends, or what triggered the burst. It has been
suggested that in their quiescent phase HII galaxies may be related to
dwarf irregulars (dI) or dwarf elliptical galaxies (dE).  Bothun \etal
(1986) made a comparative study of dIs and dEs in the Virgo cluster
based on the colour distributions and structural properties derived
from exponential fits to the surface brightness profiles (e.g. scale
length and central brightness).  They conclude that dIs are not
progenitors of dEs, but they seem to form a parallel sequence of dwarf
galaxies.  The fading of dIs would make them very diffuse and place
them below the detection threshold of photographic plates.  They
propose that Blue Compact Galaxies (BCG's, of which HII galaxies are a
subset) could probably be gas-rich analog of dEs.  Meurer, Mackie \&
Carignan (1994) have studied the structural properties of the dwarf
amorphous galaxy NGC 2915 and compared with the properties of NGC 1705
(Meurer, Freeman \& Dopita 1992) and NGC 5253 from the work of and
S\'ersic \& Donzelli (1992).  They find that their luminosity profiles
show two components indicating the presence of two distinct stellar
populations.  The inner component represents the fraction of the
galaxy dominated by hydrogen gas photoionized by the embedded massive
star clusters. Its (B-R) colour profile is increasingly bluer inwards.
The outer component has an exponential luminosity (also found for dE's
and dIrr's) and a constant redder colour likely to represent an old
stellar population remnant from a previous burst of star formation.
Their main conclusion is that these galaxies are nearby BCG's that may
provide a better insight on the properties of this type of galaxy and
their connection with other dwarf galaxies.  Kunth, Maurogordato \&
Vigroux (1988) analysed a small sample of BCG's to derive surface
brightness profiles by ellipse fitting to different isophotal levels.
Their results show that the BCG's present a \,``mixed bag of
morphologies\,''.  They find that the outer parts of the galaxies can
be best fitted by a power law compared with those of elliptical
galaxies.

No definitive study has been made on the morphology of HII galaxies
(or BCG's for that matter) up to now.  The previous attempts have
shown an extensive range of shapes from the most compact and
apparently isolated to some clearly revealing diffuse extensions,
multiple tails, and visually merging systems (Loose \& Thuan 1985;
Melnick 1987; Kunth, Maurogordato \& Vigroux 1988; Salzer, MacAlpine
\& Boroson 1989b).  Loose \& Thuan (1985) have devised a
classification scheme based on the shape and location of the burst in
relation with the whole optical structure and the shapes of the outer
envelopes.  Melnick (1987) describes the systems in terms of being
interacting, multiple, or isolated.  He has preliminarily reported
that 50\% of the HII galaxies in his sample are star-like and
isolated.  Salzer, MacAlpine \& Boroson (1989b), on the other hand,
adopted a more detailed classification scheme.  Primarily based on the
absolute magnitude, size, and morphology, with some spectroscopic
information as a secondary indicator, they identified 10 different
classes of objects for a sample of emission line objects from the
University of Michigan (UM) objective prism survey.  Their sample
includes some Seyfert galaxies as well as interacting pairs of disk
galaxies, starburst nuclei and giant irregulars.  Most HII galaxies in
our sample are classified as \,``dwarf HII hot spot galaxies\,'',
\,``HII hot spot galaxies\,'' or \,``Sargent \& Searle objects\,''.

We have used surface photometry in order to study the morphology and
structural properties of HII galaxies.  In Section~\ref{morph:sample},
we first present the data sample.  In \S\,\ref{morph:results}, we
present our results based on the analysis of the CCD images, as well
as structural aspects based on the luminosity profiles. In
\S\,\ref{morph:discussion}, we discuss the results, and finally in
\S\,\ref{morph:conclusions} we present some  conclusions.

\section{The Sample}\label{morph:sample}

The objects in the present study were selected from the sub-sample of
the {\em Spectrophotometric Catalogue of HII Galaxies} (Terlevich
\etal 1991, hereafter SCHG) used by Melnick, Terlevich \& Moles (1988)
in the application of HII galaxies as distance indicators.  It
consists of 39 HII galaxies from SCHG brighter than F(H$\beta$)$= 5
\times 10^{-15}$ \ergsqcmsec~and with H$\beta$ equivalent widths,
W(H$\beta$), larger than 30 \AA.  The brightness criterion was adopted
in order to facilitate the determination of emission line profiles,
while objects with large W(H$\beta$) were selected to minimize age
effects and contamination from an underlying stellar population.
Thus, these selection criteria yield a more homogeneous sub-sample of
the SCHG, namely, younger starbursts.  We have reproduced the relevant
data of Melnick, Terlevich \& Moles (1988) in columns 1-8 in
Table~\ref{tab:morph}.  Columns 9-12 give the results of our
morphological analysis.  Column 9 (prof.) describes the profile type
defined in Section~\ref{discuss:profiles}.  Column 10 (mult.)
describes whether the object shows a single (single), two (double), or
more than two (multiple) knots of star formation. Column 11 (ext.)
indicates whether the galaxy shows extensions or signs of distorted
outer isophotes which is the primary criterion of the morphological
classification given in Column 12 (type) of Table~\ref{tab:morph}, and
described in detail in Section~\ref{morph:class}.  A colon next to a
value in the table indicates that the classification is uncertain.

\begin{table*}
\centering
\caption{Spectroscopic data and the morphological description:
(1) name as in SCHG; (2) other name; (3) redshift; (4) line width
($\sigma$) in \kmsec; (5) H$\beta$ flux (FH$\beta$) in units of
$10^{-13}$ \ergsqcmsec; (6) extinction coefficient (CH$\beta$); (7)
H$\beta$ equivalent width in \AA~ (WH$\beta$) and (8) the oxygen
abundance in units of 12+log(O/H).  Columns (9)-(12) give the
morphological description and are discussed in the text.}

\begin{tabular}{|c|l|c|r|l|l|r|l|c|c|c|l|} \hline
SCHG & \multicolumn{1}{|c|}{{\em other name}} &  redshift & \multicolumn{1}{|c|}{$\sigma$} & FH$\beta$&
CH$\beta$ & WH$\beta$ & O/H & prof.& mult. & ext.&\multicolumn{1}{|c|}{type}\\ 
(1) & \multicolumn{1}{|c|}{(2)} &
\multicolumn{1}{|c|}{(3)}&\multicolumn{1}{|c|}{(4)} &
\multicolumn{1}{|c|}{(5)}&\multicolumn{1}{|c|}{(6)} &
\multicolumn{1}{|c|}{(7)}&\multicolumn{1}{|c|}{(8)} &
\multicolumn{1}{|c|}{(9)}&\multicolumn{1}{|c|}{(10)} &
\multicolumn{1}{|c|}{(11)}& \multicolumn{1}{|c|}{(12)} \\ \hline
0341--407 & Cam0341--4045E& 0.0147&23.0&0.44&0.28&140&8.04 &dd &multiple&$\surd$ & I  \\
`` & Cam0341--4045W& 0.0147&23.0&0.23&0.32&40 &  &&&&  \\
 ------ & Cam0357--3915 & 0.0741&51.1&0.47&0.17&180&7.87& d &single& &  II \\
 ------  & Cam08--28A    & 0.0537&49.1&0.28&0.77&35 &8.40& dd&multiple& $\surd$  & I   \\
 ------  & Cam0840+1044 & 0.0115&34.0&0.10&0.39&55 &&bd&single& $\surd$ & II:  \\
 ------  & Cam0840+1201 & 0.0305&36.5&0.48&0.52&105&7.88& d &double& $\surd$  & I  \\
 ------  & Cam1148--2020 & 0.0119&33.3&1.80&0.30&230&8.01& dd&multiple& $\surd$  &I   \\
 ------  & Cam12--39     & 0.0667&84.5&0.24&4.88&200&8.20& dd& double& & II  \\
1053+064 & Fairall 30   & 0.0035&21.8&1.80&0.22&90 &8.01& bd& single& &   II  \\
1042+097 & Fairall 2    & 0.0556&38.8&0.27&0.40&100&8.11& d& single & & I:  \\
 ------  & Cam1212+1158 & 0.0228&34.2&0.13&0.00   &60 && d&single& &    II \\
0104--388 & Tol0104--388   & 0.0211&49.0&0.30 &0.39&50 &&bd& single&&  II    \\
0127--397 & Tol0127--397   & 0.0160&33.7&0.50&0.51&40 & &d&single& &  II  \\
0226--390 & Tol0226--390   & 0.0484&89.9&0.57&0.56&90 &8.42&bd&single& $\surd$  &   I    \\
0242--387 & Tol0242--387   & 0.1260&134.0 & 0.22&  0.79&  60&8.23& d&single& $\surd$&  I  \\
0440--381 & Tol0440--381   & 0.0412&39.7&0.30 &0.32&35 &8.31& d& single & $\surd$& I \\
0513--393 & Tol0513--393   & 0.0502&33.2&0.18&0.29&145&7.90& d&single& &   II \\
0633--415 & Tol0633--415   & 0.0177&31.8&0.45&0.35&90 &8.09& dd& multiple& $\surd$&I  \\
0645--376 & Tol0645--376   & 0.0260&32.1&0.20&0.52&50 &8.19& bd& single& $\surd$  &  I   \\
1004--296 & Tol1004--294S  & 0.0038&30.6&2.70&0.69&60 &8.23& dd&double&   &   II: \\
    ``    & Tol1004--294N  & 0.0038&32.3&3.10&0.77&60 &8.28&&& & \\
1008--287 & Tol1008--286   & 0.0141&24.0&0.30&0.52&125&8.16& bd:&  multiple& $\surd$&I  \\
1025--284 & Tol1025--284   & 0.0321&25.2&0.30&0.48&60 &8.06& dd&double&$\surd$ &  I:   \\
1116--326 & Tol1116--325   & 0.0021&12.0&0.43&0.43&275&8.31& d& single& & II  \\
1147--283 & Tol1147--283   & 0.0064&18.9&0.24&0.47&45 &7.90& d &multiple:& & II \\
1214--277 & Tol1214--277   & 0.0257&27.6&0.30&0.29&230&7.58& bd&single& $\surd$  &  I  \\
1324--276 & Tol1324--276   & 0.0064&33.4&0.87&0.35&115&8.20& dd& multiple& $\surd$  & II:   \\
1334--326 & Tol1334--326   & 0.0125&16.4&0.30 &0.22&265&8.00& dd&multiple& $\surd$  &  I  \\
1345--421 & Tol1345--420   & 0.0082&21.6&0.38&0.26&70 &8.07&  d:&single&& II   \\
 ------   & Tol1406--174   & 0.0338&23.9&0.10&0.36&50 && d&single& &  II \\
1924--416 & Tol1924--416   & 0.0093&29.9&3.83&0.17&100&7.90& dd& double& $\surd$ &   I  \\
2138--405 & Tol2138--405   & 0.0578&55.5&0.33&0.25&120&7.71&dd&double& $\surd$  &I   \\
2326--405 & Tol2326--405   & 0.0505&34.5&0.14&0.22&80 &8.03& bd&double& $\surd$  & I  \\
0142+046 & UM133        & 0.0094&17.2&0.32&0.43&65 &7.64& bd:&multiple&$\surd$ & II:  \\
0131+007 & UM336        & 0.0197&16.7&0.08&0.06&40 && bd &single&   &   II \\
1134+010 & UM439E       & 0.0039&19.7&0.48&0.60&60 &8.05&  dd&multiple&  & II: \\
1139+006 & UM448        & 0.0182&40.8&2.20&0.79&45 &8.36&  bd&double& $\surd$  &  I  \\
1147--002 & UM455        & 0.0124&20.6&0.09&0.26&55 &7.84&  d &single&   &  II  \\
1148--020 & UM461A       & 0.0031&14.5&0.70&0.40&155&7.74&dd&double&   &  II  \\
1150--021 & UM462A       & 0.0031&18.5&0.75&0.38&75 &7.98&  dd&double&   &  II \\
    ``    & UM462B       & 0.0031&18.9&1.30&0.38&90 &7.79&&&&\\
0553+034 & IIZw40       & 0.0028&35.2&1.90 &1.00&170&8.13& bd&double& $\surd$ & I \\ \hline
\end{tabular}

\label{tab:morph}
\end{table*}


\subsection{CCD images}

CCD images through a broad-band V filter ($\lambda =$ 5480 \AA,
$\Delta\lambda \approx$ 900 \AA) have been obtained with the Danish
1.54-m telescope at the European Southern Observatory (ESO) in La
Silla, Chile.  The observations were made in the period 1986-1988.
The detector was a thinned RCA CCD 512$\times$320 (0.47${\tt ''}$ per
pixel) giving a total field of $4{\tt '}\times2.5{\tt '}$.  Exposures
times were typically 1200s.  Since our main aim was the study of the
morphology, photometric calibration was not performed.  We have
nevertheless estimated magnitude zero points when data were available
in the literature. Furthermore, we have not performed any correction
for galactic extinction or reddening.  Basic data reduction consisted
of bias subtraction and flat field division, cosmic rays elimination,
and sky subtraction.

\section{Results}\label{morph:results}

\begin{figure*}

\caption[Basic morphological and spectroscopic information for 39 HII
galaxies in the present sample.]{Basic morphological and spectroscopic
illustration for 39 HII galaxies in the present sample. Sets of panels
for two objects are shown in each page.  For each object we show
typically 4 panels: From top left to bottom right: 1) position,
redshift and compiled photometric information from NED and references
therein.  When available, we also include additional photometric
information from Salzer, MacAlpine \& Boroson (1989a)(indicated in the
figure by $\ddagger$), Mazzarela \& Boroson (1993)(indicated by
$\amalg$) and Telles \& Terlevich (1996)(indicated by $\dagger$).  2) scanned
photographs of the CCD images.  Orientation is shown for each image.
The angular and physical scales are shown on each photograph. 3)
original observed spectrum of the star forming region from the
SCHG. 4) circularly averaged surface brightness profiles, centered on
the main knot component.  The magnitude zero points were determined
from differential calibration using published photometry.  They have
not been derived from absolute calibration of the present work, thus
it should be taken as an indicative surface brightness level only, and
with an estimated error of at least 0.5 mag.  No correction for
galactic extinction or reddening has been applied.  Error bars in the
profiles are 1\% sky subtraction errors.}
\label{morph:panels}
\end{figure*}

In Figure~\ref{morph:panels} we present 4 panels for each object in
the present study. Each page shows these panels for two objects. From
top left to bottom right, the figure shows: 1) position, redshift and
compiled photometric information from the NASA/IPAC Extragalactic
Database (NED) plus additional photometric information from Salzer,
MacAlpine \& Boroson (1989a)(indicated in Figure~\ref{morph:panels} by
$\ddagger$), Mazzarela \& Boroson (1993)(indicated by $\amalg$) and
Telles \& Terlevich (1996)(indicated by $\dagger$).  2) grey scale
reproduction of the CCD images. Orientation is shown for each image.
The angular and physical scales are also shown on each photograph. 3)
original illustrative observed spectrum of each star forming region
from the SCHG. 4) circularly averaged surface brightness profiles
centered on the main knot component.  The magnitude zero points were
determined from differential calibration using published photometry,
thus they should be taken as an indicative surface brightness level
only with an estimated error of at least 0.5 mag.  Error bars in the
profiles are 1\% sky subtraction errors.

\subsection{Notes on individual objects}

A schematic classification of the morphology of the HII galaxies is given
in the last four columns of  Table~\ref{tab:morph}  based on two main
criteria:

\begin{description}
\item[{\em multiplicity}:] describes whether the HII galaxy has one (single)
dominant giant HII region, two (double), or more than two main knots of
star formation (multiple).

\vspace{2mm}

\item[{\em outer structure}:] denotes the presence or absence of distorted,
irregular extensions, fans or tails beyond the star forming regions.
\end{description}

The results of these two morphological description criteria are given
in columns 10 (mult) and 11 (ext) in Table~\ref{tab:morph},
respectively.  Some additional notes on individual objects are given
below:

\begin{description}

\item [SCHG 0341--407] [Cam 0341--4045] Double system; signs of multiplicity
can be seen in the West knot of star formation. The faint extension on
the NW direction may be all part of the same tail-like structure.
This object has signs of extensions and possible merger.

\vspace{2mm}

\item [Cam 0357--3915] Single; very compact; stellar object. 

\vspace{2mm}

\item [Cam 08--28] This Markarian object of high luminosity (M$_{\rm V}
\sim -22$) is in no sense a dwarf system.  It has multiple structure
and very irregular shape with clear signs of tails of faint surface
brightness emanating from the main body to the North and South; one of
the extreme examples in this sample of a possible merging system.

\vspace{2mm}

\item [Cam 0840+1044] Faint extension  in the E direction but no sign
of interaction. This object is 2 arcminutes South from a bright
foreground SBdm galaxy and close to a bright star in E.

\vspace{2mm}

\item [Cam 0840+1200] L-shaped outer isophotes; irregular inner
structures possibly due to multiple regions of star formation; likely
to be product of a merger.

\vspace{2mm}

\item [Cam 1148--2020] Multiple; main burst knot lies in the center
surrounded by a ring of 3 or 4 other regions; extended outer
isophotes.

\vspace{2mm}

\item [Cam12--39] Compact double system; no signs  of extensions.

\vspace{2mm}

\item [SCHG 1053+064] [Fairall 30] Intrinsically very compact single system at low
redshift with no sign of extensions; very close to bright star to West.

\vspace{2mm}

\item [SCHG 1042+097] [Fairall 2] Possible unresolved double system; indication of
outer faint irregular extension to W and E. It would probably show
more disturbed morphology if it were at a lower redshift.

\vspace{2mm}

\item [Cam1212+1158]  Single; regular isophotes.

\vspace{2mm}

\item [SCHG 0104--388] [Tololo 0104--388] Single and compact; close to a foreground
spiral galaxy.

\addtocounter{page}{20}

\vspace{2mm}

\item [SCHG 0127--397] [Tololo 0127--397] Single and compact.

\vspace{2mm}

\item [SCHG 0226--390] [Tololo 0226--390] Single; faint extensions from the center to
the East.

\vspace{2mm}

\item [SCHG 0242--387] [Tololo 0242--387]  Sign of a possible tail (more visible in the
CCD image) to the East at very faint surface brightness level.  This
object is the highest redshift galaxy in the sample.  It is a member of
a small group of galaxies at Z=0.126.

\vspace{2mm}

\item [SCHG 0440--381] [Tololo 0440--381] Single knot; possibly unresolved double
system;  faint blob emanating at the South-East direction.

\vspace{2mm}

\item [SCHG 0513--393] [Tololo 0513--393] Single object (southern most object in this
photograph); regular shape;  no sign of interaction.

\vspace{2mm}

\item [SCHG 0633--415] [Tololo 0633--415] Spectacular peculiar galaxy at low redshift;
very irregular morphology; bright; streamer westwards; interacting
system. Most of the [OIII] emission comes from the main knot in the
eastern component.  Very faint line emission is also seen the western
component in the narrow band [OIII] image.

\vspace{2mm}

\item [SCHG 0645--376] [Tololo 0645--376] The main body is of regular shape with its
burst located in the center, but faint extensions are seen to the East
and to the West.

\vspace{2mm}

\item [SCHG 1004--296] [Tololo 1004--294] Bright low redshift galaxy in cluster; regular
outer isophotes; double knots; amorphous galaxy.

\vspace{2mm}

\item [SCHG 1008--287] [Tololo 1008--286] Irregular; interacting system; two
main regions of star formation embedded in common irregular envelope
and bridged by faint extensions as revealed in the [OIII] image.

\vspace{2mm}

\item [SCHG 1025--284] [Tololo 1025--284] Double system embedded in common envelope.

\vspace{2mm}

\item [SCHG 1116--326] [Tololo 1116--325] Very compact object at low redshift. 

\vspace{2mm}

\item [SCHG 1147--283] [Tololo 1147--283] Single amorphous object at low galactic
latitude;  knots to the East may be stars superposed on the galaxy
image.

\vspace{2mm}

\item [SCHG 1214--277] [Tololo 1214--277] Single knot object with extensions along the
North-South direction; possible close projected companion at 20
arcseconds to the South.

\vspace{2mm}

\item [SCHG 1324--276] [Tololo 1324--276] Multiple knots at low redshift; in cluster;
signs of faint extension along the main body;  foreground bright star
superposed on the South-East end.

\vspace{2mm}

\item [SCHG 1334--326] [Tololo 1334--326] Multiple system of 4 knots at
the centre;  the most intense knot being the South West condensation;
long fan emanating from the main body to the South direction only;
object in cluster;  bright star $\approx$ 1 arcminute to the
North-West direction. Object is at low galactic latitude.

\vspace{2mm}

\item [SCHG 1345--421] [Tololo 1345--420] Single object with regular outer isophotes
(poor image quality).

\vspace{2mm}

\item [Tololo 1406--174] Single very compact stellar object.

\vspace{2mm}

\item [SCHG 1924--416] [Tololo 1924--416]  Star on South-East; Multiple knots;
almost triangle outer isophotes;  The central region of this HII
galaxy is irregular and knotty.

\vspace{2mm}

\item [SCHG 2138--405] [Tololo 2138--405] Peculiar system in interaction; two distinct
components.  The main line emitting region is in the North while the
South component has an amorphous structure.  The line emitting region
(North) seems to be double or of multiple knots.

\vspace{2mm}

\item [SCHG 2326--405] [Tololo 2326--405] Extended cometary-like object with the main
burst at its \,``head\,'' at eastern-end of the elongated body;
bright star at SE.

\vspace{2mm}

\item [SCHG 0142+046] [UM 133] Cometary-like or Magellanic-like object with main burst
at the far South end embedded in a rather regular elongated envelope;
possible multiple knots along the apparent \,``bar\,'';  no evidence of
interaction.  This object is at low redshift. It would probably be
characterized as a single object with a fuzz if it were at a larger
distance.

\vspace{2mm}

\item [SCHG 0131+007] [UM 336] Single object with regular outer isophotes. At a
moderate redshift this object is not very compact resembling more
objects with amorphous structure.

\vspace{2mm}

\item [SCHG 0134--010] [UM 439] Low redshift multiple object. Main burst is in the South
end; faint extensions along the main body; bright star 1.5 arcminutes
to South-East.

\vspace{2mm}

\item [SCHG 1139+006] [UM 448] High spatial resolution imaging of this
object (Telles \& Terlevich 1996) has shown double internal structure
of spiral-like shape.  This is a peculiar object with a visible fan,
emanating directly from the main burst extending at least 2 arcminutes
to the South-West .

\vspace{2mm}

\item [SCHG 1147--002] [UM 455] is an elliptical shape object with the line emission
region off-center. It may be an unresolved double;  bright star
$\approx$ 1.5 arcminutes to the North-East;  in cluster;  uncatalogued
galaxy at 20 arcseconds to the East with no line emission.

\vspace{2mm}

\item [SCHG 1148--020] [UM 461] Nearby double system.  The main knot in the North is of
much larger luminosity than the secondary one.  At at moderate redshift
this object would look very compact.  The outer isophotes are regular
despite the inner double structure being off-center.

\vspace{2mm}

\item [SCHG 1150--021] [UM 462] Nearby double system.  As opposed to UM 461 which is a
close companion in the group, this object has the inner double
structure in the center in relation with the outer regular isophotes.

\vspace{2mm}

\item [SCHG 0553+034] [II Zw 40] This is a low redshift object with compact
core and fan jets (South) and (South-East).  This object has two clear
different components which may represent a merging system.  At
moderate redshift its small linear size would likely hide these
features. Despite its very low galactic latitude this HII galaxy has
been targeted for various different studies in the literature. Recent
work using HST observations by Vacca (1994) reveals that the main
burst is split into smaller ionizing regions which are unresolved in
ground-based observations.

\end{description}


\subsection{General notes on morphology}

Some additional general remarks  can be made about the present sample:

\begin{itemize}

\item  Objects described to have a single giant HII region usually 
have their burst at the center (Cam0357-3915, SCHG 1053+064,
Cam1212+1158, SCHG 0104--388, SCHG 0226--390, SCHG 0242--387, SCHG
0513--393, SCHG 0645--376, SCHG 1116--326, SCHG 1345--420, SCHG
1406--174, SCHG 0131+007).

\vspace{2mm}

\item A few objects have \,``pear-like\,'' shapes (Cam 0840+1044,
SCHG 1042+097, SCHG 0440--381, SCHG 1147--002). These have been
classified as single systems with burst off-center.  However, they can
be double systems where the main knot outshines the secondary. For
example, SCHG 1147--002 (UM 461) is a compact object with a double
burst which would resemble a
\,``pear-like\,'' object, had it been located at a larger redshift.

\vspace{2mm}

\item Systems which appear to be double at a moderate redshift may
split the two main bursts into smaller components as  in the
nearby example of SCHG 1324--276. 

\vspace{2mm}

\item None of the galaxies in this sample can be classified as a
classical \,``starburst galaxy\,'', i.e.  a nuclear burst
on an otherwise normal spiral galaxy.  Typical HII galaxies
do not show spiral structure.

\end{itemize}

\begin{figure}
\protect\centerline{
\epsfxsize=3.5in\epsffile[ 30 160 580 720]
{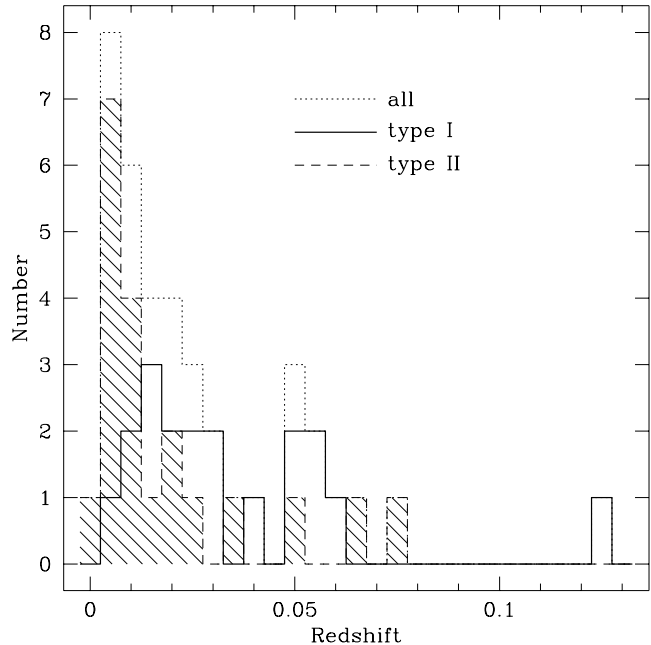}
}
        \caption{Redshift Distribution for Type I and Type II HII galaxies.
}
        \label{hist:redshift}
\end{figure}

\subsection{A classification scheme for HII galaxies}\label{morph:class}

When studying the morphological properties of HII galaxies, it is
important to bear in mind that, although all contain at least one
giant region of star formation (whether centered on the nucleus or
not), this class presents a wide variety of morphologies.  They are,
in fact, a very inhomogeneous morphological class of galaxies where
the common property is the dominant giant HII region.  The results,
presented in the last columns of Table~\ref{tab:morph} and described
in more details in the previous section reveal that the presence of an
extended component or possible underlying galaxy is evident in about
half of these objects for which CCD images have been obtained.  We
also find that some of these objects show possible features of
interacting or merging systems (wisps, tidal tails or irregular fuzzy
extensions).  More than one giant knot of star formation seems to be a
common feature, although, this seems unrelated to whether the object
is apparently interacting or isolated. About one third of the galaxies
in the sample are actually single stellar-like objects with no
evidence for extensions or fuzz.
 
On the basis of the shape of the {\it outer isophotes}, HII galaxies can be
segregated in two broad groups:

\begin{description}
\item [Type I objects] have disturbed morphologies and 
irregular outer isophotes, fuzz or tails.

\item [Type II objects] are symmetric and regular objects,
regardless of the multiplicity of the starburst region (i.e. their
internal structure).
\end{description}

An indication of the possible existence of these two morphological
types of HII galaxies was already preliminarily reported in Telles \&
Terlevich (1994).

Clearly, the perception of the morphology depends on redshift.
Individual star-forming regions within the galaxies may only be
resolved for nearby brighter objects, while morphological details of
systems at larger redshifts (z $>$ 0.02) may be smeared out rendering
the galaxies with a smoother compact appearance.  HII galaxies at
larger redshift which still show morphological details will be high
luminosity systems that may belong to a possibly more disturbed class
of starburst galaxies.  These are not dwarfs.  On the other hand,
although small morphological features would be naturally better
visible for galaxies at low redshift the distribution of redshifts
shown in figure~\ref{hist:redshift} shows that these Type II HII
galaxies are mostly low redshift galaxies.  Therefore, the fact that
we do not see faint detailed structures in these low redshift, bright,
low luminosity systems is not a resolution effect.  Objects which are
single and compact with no sign of extended envelope or tidal tails at
low redshift, such as SCHG 1053+064, Cam1212+1158, SCHG 0104--388,
SCHG 0127--397, SCHG 1116--326, SCHG 1345--421, Tol 1406--174 and SCHG
0131+007 are of particular interest.  They may be truly young galaxies
(in the sense of being experiencing their first major star formation
episode) at low redshift. Thus, these objects can provide us with an
insight on a era which was marked by the formation and early evolution
of the present luminous galaxies.

\subsection{The burst and galaxy sizes }

\begin{figure}
\protect\centerline{
\epsfxsize=3.5in\epsffile[ 30 160 580 720]{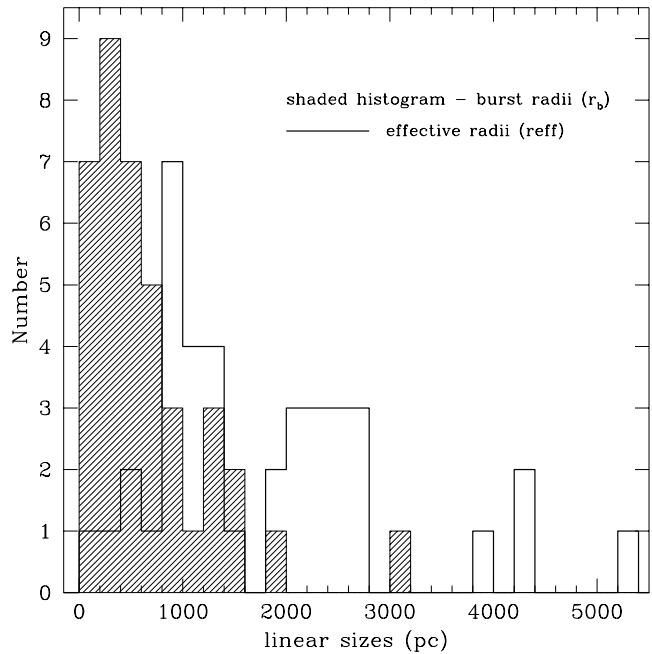}
} 
\caption[Distribution of linear radii of the bursts of star
        formation in HII galaxies.]{Distribution of linear radii of
        the bursts of star formation in HII galaxies (hatched
        histograms).  The thick line histograms show the effective
        radii of the HII galaxies in the V filter from their
        luminosity curves of growth.  } 
\label{hist:sizes}
\end{figure}

As mentioned by Djorgovski \& Davis (1987) it is difficult to define a
radial scale.  In principle we would need a radial scale independent
from the magnitude and surface brightness calibration, which means
that an isophotal radius is not ideal.  However, we can use a radial
scale derived from the surface brightness profiles, although we do not
know the exact form of these profiles for these bursts of star
formation.  Therefore, the derived relations may at best be indicative
of global rather than central properties of these galaxies.

We have measured the sizes of the star forming regions in the galaxies
as well as the half-light radius of HII galaxies from the
un-calibrated CCD images in the V band. Seeing measurements
($FWHM_{\ast}$) were obtained from Gaussian fits to the point spread
function from stars in each field and are listed in column 3 of
Table~\ref{tab:sizes}.  Burst diameters (HII in arcseconds) listed in
column 4 of Table~\ref{tab:sizes} are FWHM of the circular brightness
profiles centered on the peak intensity of the burst region of the
galaxy.  This method, although rather crude, provides us with a
systematic estimate of the {\em total} physical size of the ionizing
cluster that is adequate to the statistical approach to study the
scaling laws governing the structural properties of HII galaxies
(Telles \& Terlevich 1993).  In what follows, we have applied an
approximate seeing correction to the final burst sizes ($D_0^2 =
FWHM_{\rm HII}^2 - FWHM_{\ast}^2$). The results are shown in columns 6 and
7 in arcseconds and in parsec\footnote{H$_0 = 50$ \kmsecmeg~is used
throughout this paper.}, respectively.  It is important to bear in
mind that for barely resolved compact regions with angular sizes of
the order of the seeing disk the size estimates may represent upper
limits only.  This is illustrated in column 5 by the ratio of the
burst size measurement to the seeing ($FWHM_{\rm HII}/FWHM_{\ast}$).
Effective diameters (D$_{eff}$ in arcseconds and in parsec) in columns
8 \& 9 are the true half light diameters of the whole HII galaxy
centered on the peak luminosity (main burst) measured directly from
the synthetic circular aperture luminosity curve of growth for each
object obtained from star-free, sky subtracted frames.

Figure~\ref{hist:sizes} presents the distribution of physical sizes
(burst radii, r$_{\rm b}$) of the bursts of star formation as shaded
histograms and the distribution of effective radii (r$_{\rm eff}$) of HII
galaxies. From this, one can see that star forming regions have
typical sizes of hundreds of parsec and there is a significant cut off
above these values.  It is worth noting that these sizes are
representative of the {\em total} size of the ionizing cluster.  True
{\em core} sizes of the ionizing cluster will be far smaller.  If one
takes 30 Doradus, the most spectacular example of a nearby giant HII
region in the LMC, and compare the ratio of sizes of its ionized
region to its ionizing cluster resolved by the Hubble Space Telescope
(HST) (Walborn 1991), one would expect equivalent core sizes of the
ionizing stellar clusters of the gigantic HII regions in galaxies to
be at least 100 times smaller than their ionized regions, thus of the
order of a few parsec.  They will remain completely unresolved and, if
they are like 30 Dor, most of the ionizing luminosity will be produced
in this small region.

\subsection{Luminosity Profiles}\label{discuss:profiles}

\begin{table*}
\centering
\caption{Size measurements. $FWHM_{\ast}$ is the FWHM seeing from a
Gaussian fit to a stellar image in the CCD field.  $FWHM_{\rm HII}$
are the main burst sizes in arcseconds. D$_0$ are the burst sizes
corrected by the seeing as described in the text. D$_{eff}$ are the
half-total light diameters from the luminosity curves of growth.}

\begin{tabular}{c|l|c|c|c|c|r|r|r|}\hline
SCHG & \multicolumn{1}{|c|}{{\em other name}}  &  {\small{$FWHM_\ast$}} &  {\small{$FWHM_{\rm HII}$}} & 
{\small{$\frac{FWHM_{\rm HII}}{FWHM_{\ast}}$}} & 
\multicolumn{2}{c}{D$_0$}  &  \multicolumn{2}{c}{D$_{eff}$}  \\ \hline
 &  & (\,") & (\,") &   & (\,")& (pc)& (\,")& (pc)\\ \hline
0341--407 & Cam0341--4045E & 1.47 & 2.31 & 1.57 & 1.79   & 764 & 18.51  & 7915 \\
  ------ &  Cam0357--3915 & 1.53 & 2.03 & 1.33 & 1.34  & 2886  & 2.17  & 4670 \\
  ------  & Cam08--28A    & 1.28 & 2.69 & 2.11 & 2.37  & 3706  & 6.78 & 10595 \\
  ------  & Cam0840+1044  & 1.32 & 2.11 & 1.60 & 1.65   & 551  & 7.16  & 2395 \\
  ------  & Cam0840+1201  & 1.26 & 2.01 & 1.60 & 1.57  & 1393  & 5.18  & 4597 \\
  ------  & Cam1148--2020 & 1.64 & 2.75 & 1.68 & 2.20   & 763 & 15.64  & 5413 \\ 
  ------  & Cam12--39     & 1.51 & 1.94 & 1.29 & 1.22  & 2372  & 2.45  & 4752 \\
 1053+064 & Fairall 30    & 1.64 & 2.42 & 1.48 & 1.78   & 181  & 6.69   & 681 \\
 1042+097 & Fairall 2     & 1.32 & 2.18 & 1.65 & 1.73  & 2799  & 3.25  & 5256 \\
  ------  & Cam1212+1158  & 1.36 & 1.95 & 1.43 & 1.40   & 926  & 2.62  & 1737 \\
 0104--388 & Tol0104--388 & 1.13 & 2.05 & 1.82 & 1.71  & 1051  & 2.75  & 1688 \\
 0127--397 & Tol0127--397 & 1.63 & 2.62 & 1.61 & 2.05  &  955  & 5.18  & 2411 \\
 0226--390 & Tol0226--390 & 1.13 & 1.76 & 1.56 & 1.35  & 1899  & 2.73  & 3846 \\
 0242--387 & Tol0242--387 & 1.28 & 2.09 & 1.63 & 1.65  & 6050  & 2.30  & 8425 \\
 0440--381 & Tol0440--381 & 1.23 & 2.92 & 2.37 & 2.65  & 3178  & 3.37  & 4042 \\
 0513--393 & Tol0513--393 & 1.39 & 1.84 & 1.32 & 1.20  & 1747  & 2.17  & 3164 \\
 0633--415 & Tol0633--415 & 1.41 & 2.25 & 1.59 & 1.75   & 902  & 7.54  & 3880 \\
 0645--376 & Tol0645--376 & 1.34 & 2.36 & 1.77 & 1.95  & 1475  & 6.46  & 4888 \\
 1004--296 & Tol1004--294S & 1.19 & 2.14 & 1.79 & 1.78   & 196 & 18.37  & 2031 \\
 1008--287 & Tol1008--286 & 1.20 & 1.85 & 1.54 & 1.40   & 575 & 12.06  & 4946 \\
 1025--284 & Tol1025--284 & 1.19 & 1.89 & 1.60 & 1.48  & 1378  & 5.65  & 5278 \\
 1116--326 & Tol1116--325 & 1.43 & 2.21 & 1.55 & 1.69   & 103  & 2.49   & 152 \\
 1147--283 & Tol1147--283 & 1.39 & 3.80 & 2.73 & 3.53   & 658  & 9.70  & 1806 \\
 1214--277 & Tol1214--277 & 1.14 & 1.74 & 1.54 & 1.32   & 989  & 2.25  & 1683 \\
 1324--276 & Tol1324--276 & 1.11 & 2.16 & 1.94 & 1.85   & 344 & 13.66  & 2543 \\
 1334--326 & Tol1334--326 & 1.50 & 2.12 & 1.42 & 1.50   & 545  & 7.16  & 2603 \\
 1345--421 & Tol1345--420 & 1.58 & 3.75 & 2.38 & 3.41   & 812 & 10.93  & 2607 \\
  ------   & Tol1406--174 & 1.41 & 2.03 & 1.43 & 1.45  & 1427  & 2.15  & 2112 \\
 1924--416 & Tol1924--416 & 1.52 & 4.89 & 3.22 & 4.65  & 1258  & 6.03  & 1631 \\
 2138--405 & Tol2138--405 & 1.11 & 1.98 & 1.78 & 1.64  & 2761  & 2.57  & 4324 \\
 2326--405 & Tol2326--405 & 1.13 & 1.76 & 1.56 & 1.35  & 1981  & 2.83  & 4151 \\
 0142+046 & UM133         & 1.62 & 3.28 & 2.03 & 2.86   & 781 & 32.03  & 8758 \\
 0131+007 & UM336         & 1.10 & 1.98 & 1.79 & 1.64   & 941  & 3.81  & 2181 \\
 1134+010 & UM439         & 1.38 & 1.96 & 1.42 & 1.39   & 158 & 15.64  & 1774 \\
 1139+006 & UM448         & 1.36 & 4.94 & 3.64 & 4.75  & 2515  & 9.42  & 4987 \\
 1147--002 & UM455        & 1.75 & 2.36 & 1.35 & 1.59   & 572  & 5.12  & 1845 \\
 1148--020 & UM461A        & 1.39 & 3.52 & 2.53 & 3.23   & 291  & 9.24   & 833 \\
 1150--021 & UM462A        & 1.49 & 7.19 & 4.83 & 7.04   & 635 & 13.80  & 1244 \\
 0553+034 & IIZw40        & 1.45 & 2.76 & 1.90 & 2.35   & 191 & 12.25   & 997 \\ \hline
\end{tabular}
\label{tab:sizes}
\end{table*}


We have adopted the simplest procedure of fitting {\em circularly}
averaged radial profiles to represent the light distribution in the
isophotes of HII galaxies.  Fitting ellipses to the irregular
isophotes of HII galaxies does not produce any additional information
about the true structure of one particular galaxy.  In either case the
resulting profiles are similar.  Figure~\ref{morph:panels} shows the
derived luminosity profiles, from star-free frames, for each HII
galaxy in our present sample represented as surface magnitude $\mu(r)$
plotted against radius $r$.  In this representation an exponential
profile is a straight line.

\begin{figure}
\protect\centerline{
\epsfxsize=3.5in\epsffile[ 1 1 245 210]
{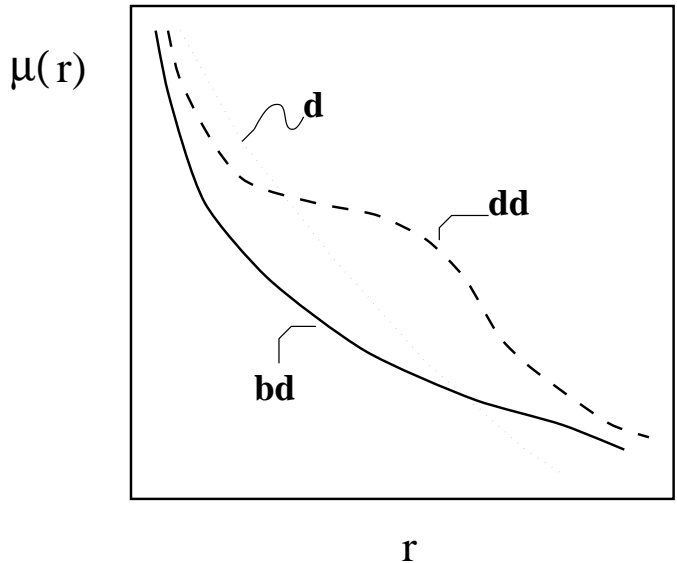}
}
\caption{Profile types.}
\label{profiles}
\end{figure}

Figure~\ref{profiles} shows the three main types of overall light
profiles found among HII galaxies.  The power law profile (solid line)
is mostly found in galaxies with long extensions.  These resemble,
somewhat, the two-component profiles such as bulge + disk profiles
typical of bright early type spiral galaxies. They trace light to
larger angular sizes [e.g. SCHG 0633--415, SCHG 2326--405, SCHG
0553+034 (II Zw 40)]. Exponential profiles (dotted lines) typically
describe the more compact, small angular size objects and are heavily
affected by the point spread function (e.g. Cam 0357--3915, SCHG
1042+097, SCHG 1116--326).  Platform profiles (dashed line) are found
for galaxies with more than one main knot [e.g.  SCHG 0341--407, SCHG
1004--296 and SCHG 1134+010 (UM 439)].

We will label the three typical types of profiles as follows:
\begin{description}
\item[{\bf d}] A single exponential fit represents well the whole range of
radii of the profile.
\item[{\bf dd}] Double profile with a \,``platform\,'' due to the double
morphology.  An exponential law is well fitted to the outer regions.
\item[{\bf bd}] a steep bright central region and outer disk-like component.
No attempt is made to perform a bulge-disk like decomposition because
of the unknown relation of the light in the burst region with the mass
density and the unknown analytical form of the light profile in the
inner region.  The exponential fit represents well the outer component
only.
\end{description}
The results of this classification scheme for the light profiles of
HII galaxies are shown in column 9 of Table~\ref{tab:morph}.

A natural further step in the analysis of the luminosity profiles of
HII galaxies is to fit the profiles with known scaling laws.  The most
common scaling laws are: exponential $\mu(r) \propto r$ (Freeman
1970); $\mu(r) \propto r^{1/4}$ (de Vaucouleurs 1948); $\mu(r) \propto
log(r)$ (Bahcall 1977); $I(r) = I_0 (1+\frac{r^2}{r_0^2})^{-1}$
(Hubble 1930).  Although these photometric laws describe the
luminosity profiles of normal galaxies they should actually not be
used as morphology descriptors; an exponential profile does not imply
a disky system unless a disk is clearly seen or there is other
indication that this may be the case. Surface photometry alone is
helpful but not sufficient to determine the spatial structure of a
particular dwarf whether it may be spheroidal or disk-like.  We have
studied which of these common scaling laws best represents the overall
shape of each HII galaxy.  The main conclusion from this study is that
{\em outer parts} of all three types of luminosity profiles of HII
galaxies are well described by an exponential scaling law
(i.e. exponential fits have the lowest standard deviation when the outer
profiles are fitted to the same radial range).  These findings for HII
galaxies agree with those of Vader \& Chaboyer (1994) who find a
general predominance of exponential profiles among dwarf of various
types although composite profile and early type r$^{1/4}$ similar to
giant ellipticals do occur in some cases.

\begin{table}
\centering
\caption{Structural parameters from exponential fits:  $\mu_0$ is the
extrapolated central surface brightness and $r_0$ is the scale length
(in arcseconds) of an exponential fit to the {\em extensions} of the
luminosity profiles.  }

\begin{tabular}{clcr} \hline
SCHG & \multicolumn{1}{c}{{\em other name}} & $\mu_0$ & $r_0$\\ \hline
0341--407 & Cam0341--4045 &          &       10.4  \\
  ------ &  Cam0357--3915 &          &        0.6  \\ 
  ------  & Cam08--28A    &  22.1   &        4.9  \\ 
  ------  & Cam0840+1044  &   23.2   &        3.4  \\ 
  ------  & Cam0840+1201  &          &        1.6  \\ 
  ------  & Cam1148--2020 &         &        7.4  \\ 
  ------  & Cam12--39     &         &        1.3  \\ 
 1053+064 & Fairall 30    &          &        2.4  \\ 
 1042+097 & Fairall 2     &          &        1.0  \\ 
  ------  & Cam1212+1158  &   19.8   &        1.0  \\ 
 0104--388 & Tol0104--388  &        &        1.7  \\ 
 0127--397 & Tol0127--397  &        &        3.1  \\ 
 0226--390 & Tol0226--390  &        &        1.8  \\ 
 0242--387 & Tol0242--387  &        &        0.9  \\ 
 0440--381 & Tol0440--381  &        &        1.1  \\ 
 0513--393 & Tol0513--393  &        &        1.0  \\ 
 0633--415 & Tol0633--415  & 25.2   &       15.0  \\ 
 0645--376 & Tol0645--376  &        &        5.6  \\ 
 1004--296 & Tol1004--294  & 19.9   &        7.5  \\ 
 1008--287 & Tol1008--286  &        &        3.8  \\ 
 1025--284 & Tol1025--284  &        &        1.8  \\ 
 1116--326 & Tol1116--325  &        &        0.7  \\ 
 1147--283 & Tol1147--283  & 21.4   &        4.1  \\ 
 1214--277 & Tol1214--277  &        &        3.4  \\ 
 1324--276 & Tol1324--276  & 20.2   &        5.4  \\ 
 1334--326 & Tol1334--326  & 19.3   &        2.0  \\ 
 1345--421 & Tol1345--420  &        &        2.2  \\ 
  ------   & Tol1406--174  &        &        0.5  \\ 
 1924--416 & Tol1924--416  & 17.3   &        5.9  \\ 
 2138--405 & Tol2138--405  &        &        1.4  \\ 
 2326--405 & Tol2326--405  &        &        4.9  \\ 
 0142+046 & UM133         &   23.1   &       14.1  \\ 
 0131+007 & UM336         &   22.6   &        2.4  \\ 
 1134+010 & UM439         &    19.5   &        3.9  \\ 
 1139+006 & UM448         &   23.7   &       14.0  \\ 
 1147--002 & UM455        &  20.3   &        1.8  \\ 
 1148--020 & UM461        &   20.8   &        3.0  \\ 
 1150--021 & UM462        &  19.5   &        3.8  \\ 
 0553+034 & IIZw40        &   22.2   &       10.0  \\ \hline
\end{tabular}
\label{fits:results}
\end{table}

Therefore, the structural parameters derived from the exponential fits
(i.e. scale length $r_0$ and central surface brightness $\mu_0$) to
the extensions may well represent the structural properties of the
underlying galaxy in HII galaxies.  In Table~\ref{fits:results} we
present the results of such exponential fits for our present sample.
Unfortunately, we do not have absolute calibration and we are able to
give approximate central surface brightness values for only $\sim$ 40
\% of the present sample.  We also caution that no reddening
correction has been applied.  One may wish to compare the results of
the structural parameters of HII galaxies directly with the properties
of other known types of dwarfs such as dE's, dIrr's or other low
surface brightness galaxies.  A comparison may help give us some
insight on the origin of the HII galaxies and the relation among all
dwarf galaxies. However, the large uncertainty in the zero point of
the few galaxies for which we have differential calibration prevent us
from performing such comparison until more an better quality data are
available.



\begin{figure*}
\protect\centerline{
\epsfxsize=5.0in\epsffile[ 30 160 580 720]
{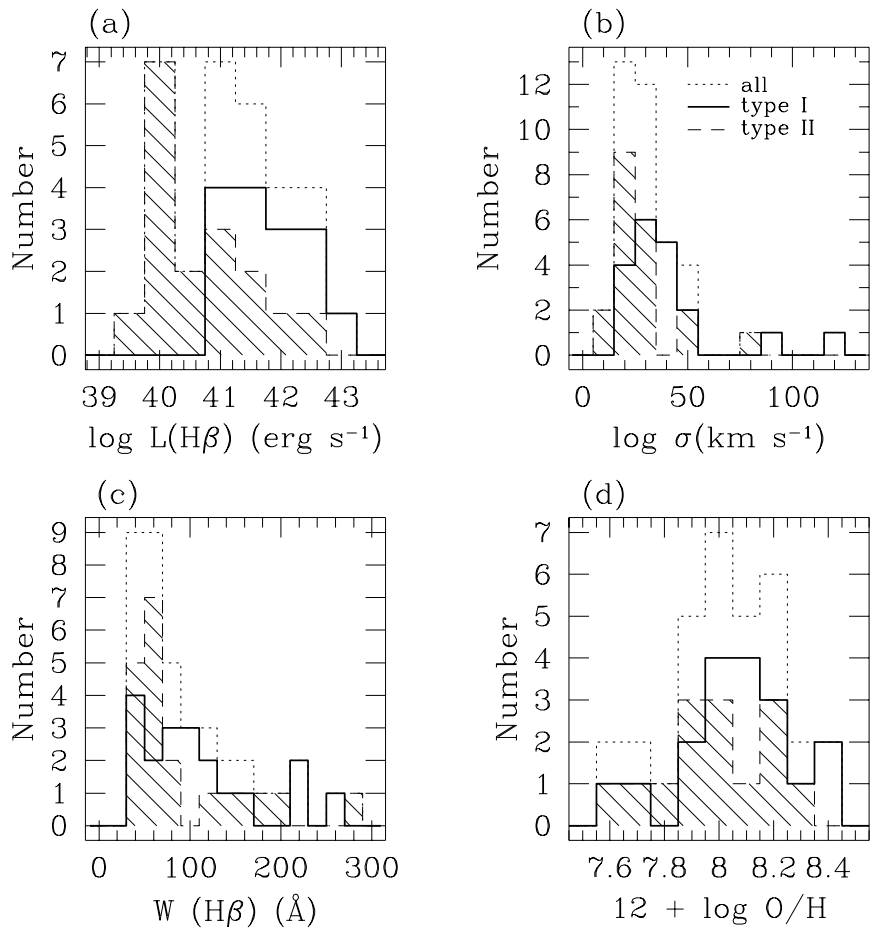}
}
\caption[Distributions of the intrinsic properties of
Type I HII galaxies and Type II HII galaxies.]{Histogram of the
spectroscopic properties segregated in Type I HII galaxies and Type II
HII galaxies. Dotted lines show the distributions for the whole
sample. Solid line histograms are Type I HII galaxies, while
short-dashed hatched histograms show Type II HII galaxies.}
\label{hist:spec}
\end{figure*}

\begin{table*}
\centering
\caption{Statistical results for the distributions of figure~5a-d.  The
mean and median of the distribution are given for Type I and Type II
HII galaxies separately. The last column gives the probability P that
the distributions of these spectroscopic properties are drawn from the
same distribution function (see text for details).  }
\begin{tabular}{|l|cc|cc|c|}\hline
	& \multicolumn{2}{c}{Type I}&\multicolumn{2}{c}{Type II}&\\ \hline
	& Mean&Median &Mean&Median&K-S test\\ \hline
$\log L(H\beta)$ (erg s$^{-1}$)&41.67$\pm$0.60&41.41$\pm$0.53&40.64$\pm$0.84
&40.30$\pm$0.48&P = 0.001\\
$\sigma$ (\kmsec)&40.0$\pm$24.2&32.9$\pm$6.6&28.7$\pm$15.6&23.1$\pm$7.3&
P = 0.02\\
W(H$\beta$) (\AA) &112$\pm$68&100$\pm$40&94$\pm$64&65$\pm$45&P = 0.18\\
$12 + \log({\rm O/H})$&8.07$\pm$0.21 &8.09$\pm$0.09 &7.99$\pm$0.19&8.01
$\pm$0.16&P = 0.51\\ \hline
\end{tabular}
\label{morph:tab_dist}
\end{table*}

\section{Discussion}\label{morph:discussion}

Figure~\ref{hist:spec} shows the distributions of the intrinsic 
properties of HII galaxies segregated in  Type I HII galaxies (solid
line histograms) and Type II HII galaxies (shaded histograms).  Also
shown, as dotted lines, are the histograms for all galaxies in the
sample.  
Table~\ref{morph:tab_dist} summarizes the main statistical parameters
from these distributions.  We have checked the trends shown in
figure~\ref{hist:spec} by performing a Kolmogorov-Smirnov test to the
unbinned distributions in order to assess statistically the
differences of the spectroscopic data for Type I and Type II HII
galaxies.  The low values of the significance level ($P \leq 0.02$)
disproves that the distributions of L(H$\beta$) and $\sigma$ can be
drawn from the same distribution function.  However, the high values
of the significance level ($P > 0.1$) for the distributions of
W(H$\beta $) and O/H indicate that the data are consistent with them
being drawn from a single distribution function.  These results show
that HII galaxies with signs of disturbed morphologies (Type I) tend
to be more luminous than Type II objects.  It can also be seen that
Type I also tend to have larger velocity dispersion ($\sigma$) in the
line emission regions in agreement with the well known correlation
between L(H$\beta$) and $\sigma$ (Terlevich \& Melnick 1981; Melnick,
Terlevich \& Moles 1988).  The metal abundance and W(H$\beta$)
distributions seem to overlap for both types of HII galaxy.  However,
the morphological disturbances of Type I HII galaxies do not seem to
be of tidal origin.  HII galaxies are not associated with bright
companions, they are mostly isolated and tend to populate low density
environments and are weakly clustered (Telles \& Terlevich 1995;
Vilchez 1995; Rosenberg, Salzer \& Moody 1994).

It is interesting to point out that other types of dwarf galaxies have
been found to show distinct systematic structural properties at both
ends of the luminosity distribution (Binggeli \& Cameron 1991,
Binggeli 1994, Ferguson \& Binggeli 1994).  Binggeli \& Cameron (1991)
have found that the break in the systematic photometric properties in
their sample of $\sim$ 200 early-type dwarfs in the Virgo cluster lies
at M$_B \approx$ --16 to --17.  They suggested that this may be caused
by a transition in the internal kinematics of the systems from disky
or rotationally supported systems to spheroidal systems.  They also
suspect a connection with the ratio of visible to dark matter; fainter
galaxies being more dark matter dominated.  The question that then arises
is whether there is indeed a real break in physical properties of all
dwarf galaxies at that luminosity.  If so, what is the relation and
underlying causes between the break at the photometric properties for
the dwarfs in Virgo and the apparent break in the morphological
properties of HII galaxies?  In any case, our results seem to suggest
a strong relation between morphology and luminosity.

\begin{figure}
\protect\centerline{
\epsfxsize=3.4in\epsffile[ 50 210 530 670]{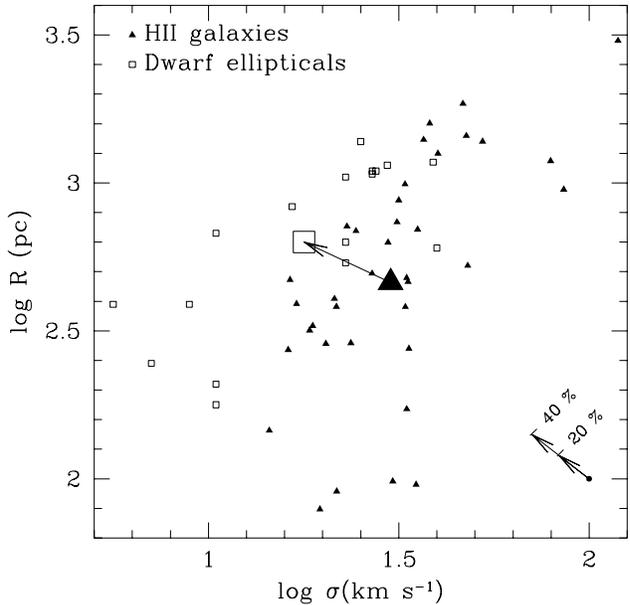}
}
\caption[The Radius vs. $\sigma$ relations for dEs and HII
galaxies.]{The Radius vs. $\sigma$ relations for dEs and HII galaxies.
R corresponds to R$_{\rm b}$ (D$_0$/2 from Table 2) for HII galaxies.
The big symbols represent the mean locus of the HII galaxies (big
solid triangle) and the mean locus of the dwarf elliptical galaxies in
the diagram (big open square). }
\label{dynam:dwarf_r}
\end{figure}

Apart from the photometric evolution of the stellar population in HII
galaxies, because of the mass loss from stellar winds and supernova
ejecta and its short dynamical time scale, dynamical evolution will
affect the system in less than a Hubble time (Terlevich 1994).  We can
predict the trend of the effect of mass loss in the dynamical
evolution of the stellar cluster.  If the removal is slow compared to
the crossing time of the system $t_{\em ff}$, then the system obeys an
adiabatic invariant (Dekel \& Silk 1986, Terlevich 1994 and references
therein), and the \,``bloating\,'' of the stellar system will be
proportional to the mass loss. For the system to remain bounded
throughout its evolution, the increase in size should be accompanied
by a corresponding decrease in velocity dispersion:

\[
\frac{r}{r_0} = \frac{{\cal M}_0}{{\cal M}} = \frac{\sigma_0}{\sigma}
\]

\noindent
where $r$ is the size of the system, ${\cal M}$ is its total mass and
$\sigma$ is the velocity dispersion.  During the evolution the total
mass lost by a \,``normal\,'' IMF population is in the range of 20\% -
40\%, considering that all ejecta leave the system.

Figure~\ref{dynam:dwarf_r} shows the [R -- $\sigma$] relation for our
sample of HII galaxies (solid triangles) and compares it with sample
of dwarf elliptical galaxies (open squares) from Peterson \& Caldwell
(1993, and references therein). The big symbols represent the mean
locus of the HII galaxies (big solid triangle) and the mean locus of
the dwarf elliptical galaxies in the diagram (big open square).  The
arrow connecting the big symbols is the observed shift from the HII
galaxies to the dwarf elliptical galaxies. This trend, namely that
dwarf ellipticals are typically slightly larger and with lower
velocity dispersion than HII galaxies, is compatible with the
expectation if the stellar population in HII galaxies evolve
dynamically into dwarf elliptical galaxies.  In addition, the
magnitude of the observed shift is in good agreement with the
estimated dynamical evolution due to mass loss by a \,``normal\,'' IMF
population ($\Delta{\cal M}/{\cal M_0} \sim 20\% - 40\%$), represented
in the figure by the arrows in the right lower corner.  The similarity
of the parametric relations of \,``aged\,'' HII galaxies with those of
other dwarf galaxies (Telles \& Terlevich 1993), together with the
indication of a possible dynamical evolution of HII galaxies into
dwarf ellipticals, may be suggestive of a close kinship among these
dwarfs.

\section{Conclusions}\label{morph:conclusions}

The morphological and structural properties of HII galaxies are
studied  in this paper.  The main conclusions of this study are:

\begin{itemize}

\item HII galaxies can be classified within two broad morphological
 types:
\begin{description}
\item[Type I] irregular systems with signs of distorted outer
isophotes, tails, wisps, fans, etc. 
\item[Type II] regular and compact systems with symmetric morphology.
\end{description}

Type I's were found to have higher luminosities and velocity
dispersions than Type II's, while the equivalent widths of H$\beta$
and oxygen abundances of the two types are roughly similar.  This
seems to indicate that the starbursts may have been triggered by
different mechanisms in the two classes of objects.


\item We find three main types of light profiles in HII galaxies. The
profile types {\em qualitatively} relate to the overall morphology of
the galaxies. The outer parts of the luminosity profiles of HII
galaxies are well represented by an exponential scaling law.  This
will allow a direct comparison of the structural parameters (scale length
and central surface brightness) with other types of dwarf galaxies and
will lead us to derive important structural properties of the
underlying systems once calibrated images are obtained.

\item While the burst sizes of HII galaxies are of the order of 
hundreds of parsec, the \,``true\,'' core radii of HII galaxies are
basically unresolved and probably only few parsecs across. Yet, most
of the ionizing luminosity produced may be coming from these very
small regions.
  
\item The similar trends of dynamically \,``aged\,'' HII galaxies, and
their relative positions in the [R -- $\sigma$] diagram support the
hypothesis of a possible evolutionary link between the two types of
galaxy.  If this is the case, dEs could be the descendants of HII
galaxies.


\end{itemize}

\section*{Acknowledgments}

ET acknowledges his grant from CNPq/Brazil.  We especially thank Richard Sword
for his art work in this paper.

\end{document}